\documentclass[10pt,conference]{IEEEtran}
\IEEEoverridecommandlockouts
\usepackage{cite}
\usepackage{amsmath,amssymb,amsfonts}
\usepackage{algorithmic}
\usepackage{graphicx}
\usepackage{textcomp}
\usepackage{xcolor}
\def\BibTeX{{\rm B\kern-.05em{\sc i\kern-.025em b}\kern-.08em
    T\kern-.1667em\lower.7ex\hbox{E}\kern-.125emX}}

\usepackage[T1]{fontenc}
\usepackage{blindtext}
\usepackage{hyperref}
\usepackage{caption}
\usepackage{subcaption}
\usepackage{xspace}
\usepackage[framemethod=tikz]{mdframed}
\definecolor{mycolor}{rgb}{0.122, 0.435, 0.698}
\newmdenv[innerlinewidth=0.5pt, roundcorner=4pt,linecolor=mycolor,innerleftmargin=6pt,
innerrightmargin=6pt,innertopmargin=6pt,innerbottommargin=6pt]{mybox}

\newcommand{\eg}{\emph{e.g.,}\xspace}

\newcommand{\etal}{\emph{et al.}\xspace}

\newcommand{\RQ}[1]{RQ$_{#1}$}

\newboolean{showedits}
\setboolean{showedits}{true} % toggle to show or hide edits
\ifthenelse{\boolean{showedits}}
{
	\newcommand{\del}[1]{\textcolor{red}{\sout{#1}}} % please delete
	\newcommand{\nbe}[3]{
		{\colorbox{#3}{\bfseries\sffamily\scriptsize\textcolor{white}{#1}}}
		{\textcolor{#3}{\sf\small$\blacktriangleright$\textit{#2}$\blacktriangleleft$}}}
}{
	\newcommand{\del}[1]{} % please delete
	
	\newcommand{\nbe}[3]{}
}

\begin{document}

\title{What is a Feature, Really? Toward a Unified Understanding Across SE Disciplines}

\author{
    \IEEEauthorblockN{Nitish Patkar\IEEEauthorrefmark{1}, 
    Aimen Fahmi\IEEEauthorrefmark{2}, 
    Timo Kehrer\IEEEauthorrefmark{3},
    Norbert Seyff\IEEEauthorrefmark{1}}
    \IEEEauthorblockA{\IEEEauthorrefmark{1}University of Applied Sciences and Arts Northwestern Switzerland (FHNW), Switzerland\\
    \{nitish.patkar, norbert.seyff\}@fhnw.ch}
    \IEEEauthorblockA{\IEEEauthorrefmark{2}University of Fribourg, Switzerland\\
    aimen.fahmi@unifr.ch}
    \IEEEauthorblockA{\IEEEauthorrefmark{3}University of Bern, Switzerland\\
    timo.kehrer@unibe.ch}
}

\maketitle

\begin{abstract}
In software engineering, the concept of a ``feature'' is widely used but inconsistently defined across disciplines such as requirements engineering (RE) and software product lines (SPL). 
This lack of consistency often results in communication gaps, rework, and inefficiencies in projects. 
To address these challenges, this paper proposes an empirical, data-driven approach to explore how features are described, implemented, and managed across real-world projects, starting with open-source software (OSS). 
By analyzing feature-related branches in OSS repositories, we identify patterns in contributor behavior, feature implementation, and project management activities. 
Our findings provide actionable insights to improve project planning, resource allocation, and team coordination. 
Additionally, we outline a roadmap to unify the understanding of features across software engineering disciplines. 
This research aims to bridge gaps between academic inquiry and practical strategies, fostering better feature planning and development workflows in diverse project environments.
\end{abstract}


%=================================%
% =========== SECTION ============%
%=================================%
\section{Introduction}
\label{sec:introduction}
In software engineering, the concept of a ``feature'' plays a central role in both research and practice. 
Features are critical units of functionality that guide product development and connect technical implementation with user needs. 
However, the definition and understanding of features vary significantly across disciplines like requirements engineering (RE), software product lines (SPL), and agile software development (ASD)~\cite{clas08,apel09,berg15,mole22}. 
This inconsistency leads to challenges such as miscommunication, rework, and inefficiencies in project management. 
These issues can result in missed deadlines, increased costs, and lower product quality.

A unified understanding of features is essential to address these challenges. 
However, there is limited empirical evidence on how features are described, implemented, and managed in real-world projects.

Existing studies in RE have not examined feature descriptions in real-world projects to assess their alignment with implementation. 
Similarly, research on implementation has mostly focused on identifying features rather than understanding them in different contexts~\cite{krug18,krug19a,krug19b}. 
These studies, while insightful, are limited to single projects and rely on manual analysis of artifacts like commit messages and pull requests.

To address this gap, We propose a data-driven approach to study the lifecycle of features across diverse projects. 
This approach involves analyzing requirements, code artifacts, and project management data. 
We will also gather insights from practitioners through surveys and interviews. 
By combining these methods, we aim to provide a unified understanding of features that benefits both researchers and practitioners.

As a first step, we examine open-source software (OSS) projects due to their transparency and availability of data. 
GitHub, a widely used platform for hosting OSS and private repositories, provides extensive data on feature development. 
By studying feature-related artifacts such as branches, issues, and pull requests, as well as contributor behavior and project activities, we aim to generate insights applicable to both OSS and proprietary software.

The proposed research is novel because it examines features across all stages of their lifecycle, from requirements to implementation and project management. 
Unlike previous studies that focused on isolated contexts, we analyze features across diverse projects and address gaps between RE, development, and project management. 
This holistic approach is especially important given the increasing complexity of modern software systems and the need to align stakeholder expectations with technical outcomes.
%For practitioners, the findings can inform data-driven estimation techniques and support resource planning and project management, ultimately helping align feature development with business goals and improving project outcomes.

In this paper, we outline our research vision, elaborate on planned contributions, and discuss their impact on the state of the art in software engineering. 
We also present preliminary results from mining GitHub.

%=================================%
% =========== SECTION ============%
%=================================%
\section{Research vision}
\label{sec:vision}
Our vision is to establish a shared understanding of the concept of a ``feature'' through empirical research to bridge gaps between software engineering disciplines and improve software development practices.
To achieve this vision, we set the following research goals:

\emph{Clarify the notion of a ``feature''} through empirical analysis of project-level data and qualitative data collected from practitioners using surveys and interviews. 
This will enable teams to align on a shared understanding of features, reducing miscommunication during development.

\emph{Examine how features are described, implemented, and managed} across different project types (\eg open-source and proprietary), sizes, and ages, and identify the impact of these factors on project outcomes. 
This will provide insight into how project context influences feature development practices, helping teams adapt processes accordingly. 

\emph{Study the complete lifecycle of features}, from initial definition to implementation, to uncover patterns and commonalities in feature development and management practices. 
By identifying recurring patterns, we can propose improvements to project planning and feature implementation strategies.

\emph{Propose strategies to reduce rework and improve feature planning}, ensuring better alignment between stakeholder expectations and technical implementation. 
By offering data-driven recommendations, we aim to improve both the quality and efficiency of feature development.

\subsection{Research questions and methodology}
There are several high-level research questions we would like to answer:
\begin{itemize}
    \item \RQ{1}: How do different stakeholders interpret and manage features throughout the development process, and how do their interactions influence project success?
    \item \RQ{2}: What patterns and challenges emerge when managing features across projects of different types and sizes?
    \item \RQ{3}: How are features typically described from a requirements engineering (RE) perspective, and what impact does the quality of these descriptions have on their successful implementation?
\end{itemize}
Our research employs a mixed-method approach, combining quantitative and qualitative techniques to gain a comprehensive understanding of how features are described, implemented, and managed in software projects. We start with empirical analysis of large-scale project data from OSS repositories to extract insights on feature development patterns and practices. 
This includes using Natural Language Processing (NLP) to identify feature types and automated analysis of project artifacts, such as branches, pull requests, and contributor activity.

In parallel, we conduct qualitative research through surveys and interviews with industry practitioners to complement our findings with real-world perspectives. 
This combination of data sources enables us to validate our observations and draw deeper insights into how feature-related practices vary across different project contexts.

Our research will expand beyond OSS to include proprietary projects and large-scale industrial systems, where we will investigate how feature management practices differ in more complex environments.

As with any empirical research, we recognize potential threats to validity. For example, the generalizability of findings from open-source projects and the limitations of automated techniques must be carefully considered. 
To address these concerns, we will refine our methods and validate our results through ongoing iterations and practitioner feedback.

\subsection{Potential research implications}
There are several implications of our research:

\emph{Bridging Disciplinary Gaps.} By providing a shared understanding of features, our findings will help align RE, SPL, and agile development practices. 
This shared perspective can reduce miscommunication within multidisciplinary teams.

\emph{SE Process and Project Management Improvements.} The data we collect will support predictive models for planning and management. 
These models will help teams estimate time-to-market, resource needs, and rework probabilities, aiding in feature prioritization and timeline management.

\emph{Automation Potential.} Our findings will lay the foundation for machine learning models that predict project variables, such as feature completion times and resource allocation. 
This will improve predictability and reduce the need for manual estimation.

\emph{Improved Collaboration.} Clearer feature descriptions and expectations will foster better alignment between technical teams and business stakeholders, leading to improved product quality and customer satisfaction.

%\subsection{Threats to the validity}
%\label{sub:threats}
%\topic{Internal Validity}: Since our approach integrates quantitative and qualitative methods, potential biases may arise from both automated analysis (\eg feature identification through branch names) and subjective interpretations during interviews or surveys. 
%To mitigate these threats, we will iteratively refine our automated techniques and use systematic coding procedures for qualitative data.

%\topic{External Validity}: Starting with open-source software (OSS) projects may limit the applicability of our findings to proprietary projects. 
%To address this, we will expand the scope to include industrial projects and diverse practitioner inputs, ensuring broader relevance and generalizability.

%\topic{Construct Validity}: Our operational definitions of features may not fully capture the concept across different SE disciplines, potentially leading to biased interpretations. 
%We address this by triangulating multiple data sources and cross-validating with practitioner perspectives.

%=================================%
% =========== SECTION ============%
%=================================%
\section{Early results from ongoing research}
\label{sec:preliminary-results}
In our ongoing research, we initiated a focused study on the use of feature branches in open-source software (OSS) projects hosted on GitHub. 
While existing studies have primarily examined GitHub issues and pull requests, branches—particularly feature branches—remain under-explored in the context of feature development. 
Given GitHub’s recommendation to use branches for implementing new features, we identified feature branches as a key artifact for analyzing how features are implemented and managed in practice. 
This preliminary investigation is an essential first step towards understanding broader patterns and practices related to our high-level research questions, particularly \RQ{1} and \RQ{2}.

Below, we present the results of our ongoing research to evaluate whether feature-related branches can serve as a reliable source for in-depth studies. 
The data collection and analysis were primarily conducted by the second author, in close collaboration with the co-authors.

\subsection{Data collection and analysis}
We used a dataset from the SEART GitHub Search platform, which includes 1,580,895 repositories~\cite{Muna17}.\footnote{https://seart-ghs.si.usi.ch/} 
The data was collected on August 8th, 2024, without any initial filters. 
To focus on active, collaborative, and popular projects, we applied the following commonly used criteria~\cite{Mci21,Kall16,Han19,Padh19}: repositories created between January 1, 2017, and December 31, 2023, with a minimum of 100 stars, at least five contributors, two or more branches, and a minimum of 100 commits. 
This filtering reduced the dataset to 53,802 repositories.

Next, we refined the dataset further by selecting only repositories with branch names following the pattern ``feature/feature-name,'' resulting in 6,637 repositories and 563,841 feature branches. 
This allowed us to distinguish between feature branches and non-feature branches for comparative analysis.

We then counted the number of feature branches per repository, the number of commits per branch, and analyzed commit activity per contributor to determine if these metrics provided sufficient data for in-depth analysis. 
Additionally, we compared commit patterns over time between feature and non-feature branches to identify distinct trends.

\subsection{Results}
\begin{figure}[h]
    \centering
    \includegraphics[width=\columnwidth]{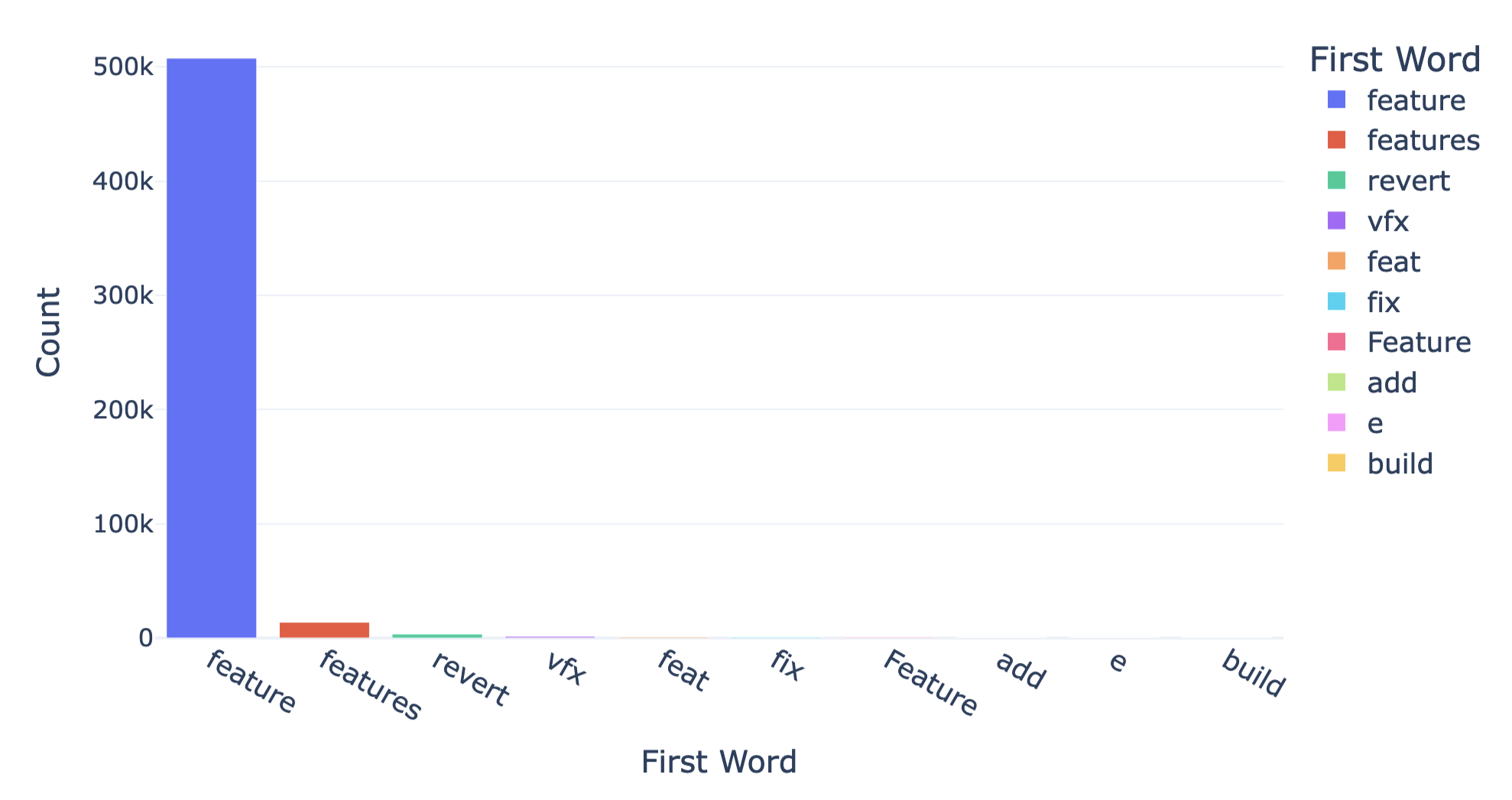}
    %\decoRule
    \caption{Top 10 most frequent first words in feature branch names}
    \label{fig:first_words}
\end{figure}

\emph{Feature development within OSS projects.} 
We found that roughly 12\% of the repositories in the SEART dataset have used feature branches or, more generally, have engaged in feature development. 
This percentage could even increase if we modify the filtering criteria and make it stricter, \eg at least 10 contributors or at least 200 stars.
Likewise, we only considered branch names containing the substring ``feat''. The actual number of branches for feature development could in fact be higher.
\autoref{fig:first_words} shows the top 10 most frequent first words found in branch names, with ``feature'' being the most common and representing 90.02\% of all branches in the dataset.

After removing outliers using the Interquartile Range (IQR) method, we found that the average number of feature branches per repository was 38.91, with a median of 21. 
We also observed that feature branches have an average of 1,690.89 commits and a median of 1,027 commits per branch. 

These initial results suggest that OSS projects do manifest feature development.
The number of feature branches per repository and the number of commits per branch indicate that feature development is prevalent and consistently used in OSS projects. 
This data sufficiency supports our intention to analyze patterns and challenges in managing features across projects (related to \RQ{2}) and lays the groundwork for understanding how feature management practices vary by project size and type.

\begin{figure}[h]
    \centering
    \includegraphics[width=\columnwidth]{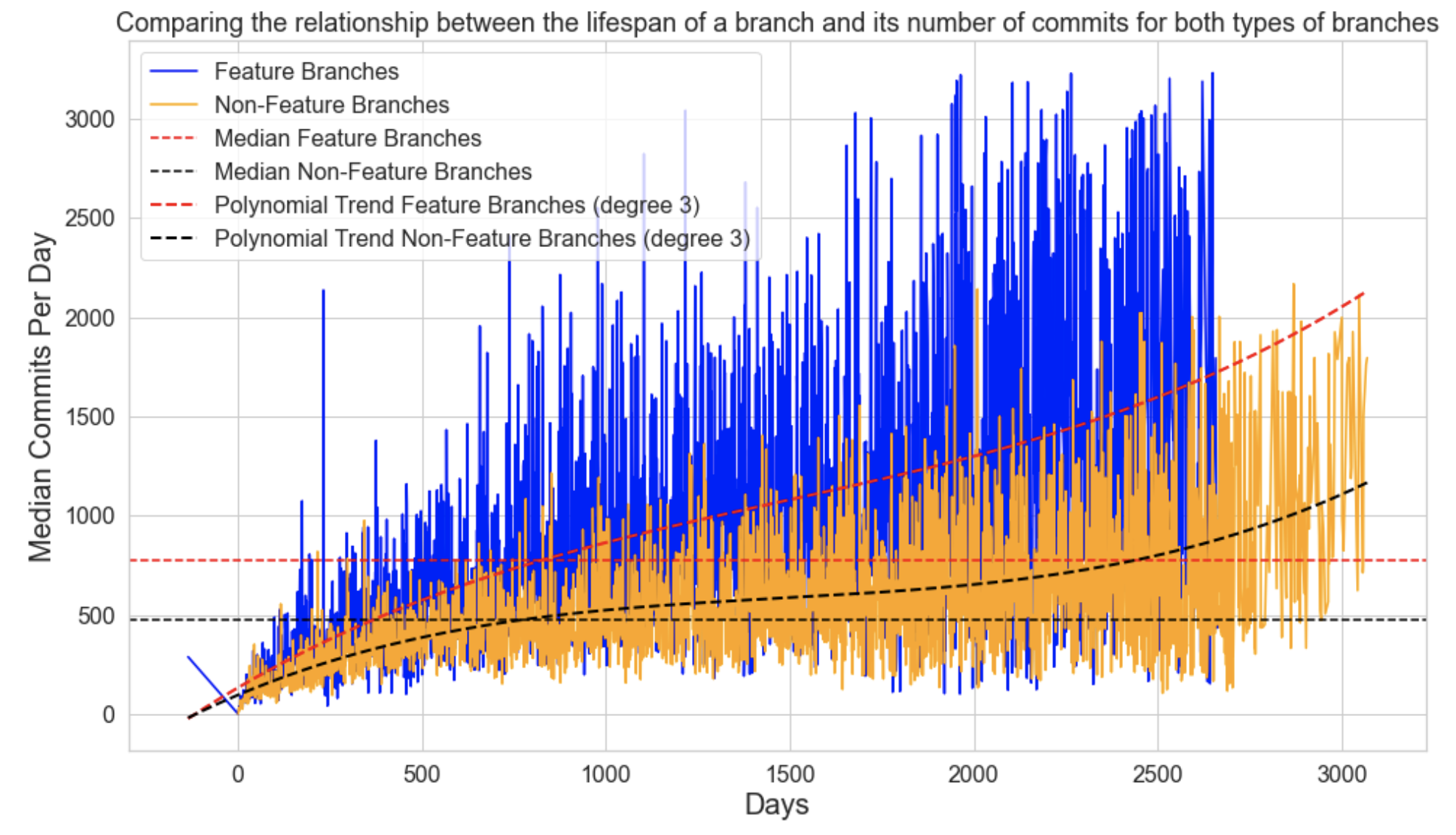}
    \caption{Commit activity over time}
    \label{fig:commit_activity_time}
\end{figure}

\emph{Feature vs. non-feature branches.}
We found that feature branches exhibit significantly higher commit activity, with a mean of 30.49 commits per contributor and a median of 24.07, compared to non-feature branches, which have a mean of 21.11 and a median of 16.44. 
An independent samples t-test yielded a t-statistic of 60.77 and a p-value of 0.0, indicating a statistically significant difference in commit activity per contributor between feature and non-feature branches.
\autoref{fig:commit_activity_time} shows the trend of median commits per day across the lifespan of both feature and non-feature branches. 
Additionally, as branch lifespan increases, median commit activity also rises, particularly for feature branches. 
A Kolmogorov-Smirnov (KS) test comparing the distributions of commit activity over time resulted in a KS-statistic of 0.305 and a p-value of 0.000, confirming a statistically significant difference in commit patterns between feature and non-feature branches.

These results indicate that developers handle feature branches distinctly, making them valuable for gaining deeper insights into contributor behavior and development patterns (related to \RQ{1}). 
The higher commit activity in feature branches suggests concentrated development efforts, where contributors frequently update and refine the codebase when working on new features. 
In-depth investigation can inform strategies for improving feature planning and reducing inefficiencies, ultimately supporting our goal of better aligning development practices with feature management strategies.

%=================================%
% =========== SECTION ============%
%=================================%
\section{Future plan}
\label{sec:future plan}
Our future research directions primarily focus on \RQ{1} and \RQ{2}, investigating how stakeholders interpret and manage features and exploring patterns in feature management across various projects. 
While our current work does not yet focus deeply on feature descriptions from a requirements engineering perspective (\RQ{3}), future expansions of this research will explore how feature descriptions impact implementation success once broader management patterns have been identified.

\emph{Analyzing Feature Branches and Issues.} We will build on our initial findings by extending the analysis of feature branches to include feature-related issues. 
This expansion will help us explore how features are planned, discussed, and tracked across different projects. 
By leveraging NLP techniques, we will identify patterns in branch naming conventions and issue descriptions. 
This will allow us to systematically analyze feature representations and understand how similar features (\eg a ``digital wallet'') are managed across diverse project types. 
These insights will reveal the interplay between high-level feature descriptions and their implementation.

\emph{Exploring Feature-Specific Pull Requests.} As the next step, we will focus on analyzing pull requests (PRs) specifically tied to feature development. 
By examining factors such as PR size, review comments, and time to merge, we aim to identify characteristics that influence the quality and success of feature development. 
This focused analysis will shed light on how code reviews and integration processes impact the implementation of complex features and reveal potential bottlenecks in feature-related workflows.

\emph{Contributor and Team Dynamics.} We will investigate the roles, behaviors, and interactions of contributors working on feature-related tasks, examining how these dynamics vary across projects of different sizes and domains. 
By comparing feature-specific contributions to more general development activities, we aim to identify the unique challenges and collaboration patterns involved in feature development. 
This will provide a richer view of how contributors’ activities impact feature management and team coordination.

\emph{Integrating Practitioner Perspectives.} In parallel, we will incorporate qualitative insights by gathering perspectives from industry practitioners through interviews and surveys. 
These insights will help us validate our findings and refine our understanding of how features are managed in real-world projects. 
Practitioner input will allow us to better contextualize our empirical results and develop guidelines aligned with both research and industry practices.

\emph{Extending to Proprietary and Large-Scale Projects.} While OSS projects provide a good starting point, our goal is to expand our analysis to include proprietary and large-scale industrial projects. This broader scope will enable us to compare feature management practices in more structured and complex environments, where features often involve stricter requirements and higher stakes. 
By doing so, we aim to develop generalizable models and strategies for effective feature planning and management across diverse settings.

%=================================%
% =========== SECTION ============%
%=================================%
\section{Related work}
\label{sec:related-work}
Many existing studies have focused on formalizing the definition of ``feature'' across different contexts, providing structured categorizations and guidelines for consistent usage~\cite{clas08,apel09, bane12,berg15,kala18,mole22}. 
They define features from perspectives such as requirements engineering, software product lines, and information science. 
However, these studies primarily address conceptual clarity rather than empirically analyzing how features are developed and managed in practice.

There also exists a vast body of research on mining GitHub data, with many studies focusing on specific aspects relevant to our work, such as contributor behavior, project management, and communication patterns. 

While Chatziasimidis \etal focus on analyzing contributor activity primarily to identify project success factors (\eg user activity and the number of collaborators)\cite{Chat15}, our research aims to understand contributor behavior specifically in the context of feature development. 
The study by Ortu \etal provides relevant insights into contributor roles and communication patterns through comment analysis in OSS projects\cite{Ortu18}. 
However, its focus is limited to comments, whereas our research aims to analyze multiple aspects of contributor behavior, including their activity in feature branches, pull requests, and issue resolution, to gain a deeper understanding of feature-specific development practices.

The study by Onoue \etal explores the diversity of developer behaviors, including coding, commenting, and issue handling patterns, but it is limited to only two OSS projects~\cite{Ono13}. 
Rahman \etal compare successful and unsuccessful pull requests across 78 GitHub projects involving 20,142 developers and 103,192 forks~\cite{Rah14}, while our analysis is broader and targeted toward understanding the lifecycle of feature development. 
Finally, Guzman \etal analyzed 60,425 commit comments from 90 projects to understand developer sentiments~\cite{Guz14}, but again, our scope is broader and more focused on feature-specific interactions.
%Borges \etal studied factors that impact the popularity of GitHub repositories~\cite{Borg16}.
%Qi \etal mined GitHub to collect diverse real-world effort data for effort estimation~\cite{Qi17}.

%Kovalenko \etal evaluated the importance of correctly handling branches when calculating file modification histories~\cite{Kova18}.

%=================================%
% =========== SECTION ============%
%=================================%
\section{Conclusion}
\label{sec:conclusion}
In this paper, we presented our vision for establishing a shared understanding of the concept of a ``feature'' in software engineering through a systematic, data-driven approach. 
By analyzing feature branches in open-source software (OSS) projects, we laid the groundwork for a broader investigation into how features are defined, implemented, and managed in real-world projects. These initial results demonstrate that feature-specific artifacts can reveal valuable insights into development patterns, contributor behavior, and project management practices.

As the next step, we will expand our analysis to include pull requests and feature-related issues to further explore how feature descriptions align with implementation and management activities. 
We also plan to integrate practitioner insights to validate our findings and refine our understanding of feature practices in diverse development contexts.

Our research has the potential to reshape how software engineering communities--from requirements engineering (RE) to software product lines (SPL) and agile teams--perceive and manage features. 
By bridging gaps between disciplines and aligning feature development and management strategies across diverse development environments, we aim to create a shared understanding that improves communication, reduces rework, and enhances project planning. 
The data we collect will enable predictive models for planning and management, forming the basis for machine learning models that forecast project variables like feature completion times and resource allocation.
With future expansions into proprietary and large-scale projects, we envision that our work will contribute not only to academic knowledge but also to practical, industry-wide transformations in feature management and planning.

%Our research opens up numerous opportunities for advancing software engineering practices, such as developing refined resource estimation techniques, establishing new methods for tracking feature evolution, and creating strategies for achieving better alignment between requirements and implementation. 

\bibliographystyle{splncs04}
\bibliography{ICSE_NIER_2025}

\end{document}